\documentclass{article}
\usepackage[left=3cm,top=2.5cm,right=3cm,bottom=2.5cm]{geometry} 

\usepackage{amsmath}
\usepackage{mathrsfs}
\usepackage{slashbox}
\usepackage{multirow}
\usepackage{epsfig}

\newcommand{\eq}{Eq.~}
\newcommand{\sect}{Sect.~}

\newcommand{\fig}{Fig.~}
\newcommand{\envelope}{(\raisebox{-.5pt}{\scalebox{1.45}{\Letter}}\kern-1.7pt)}
\newcommand{\ket}[1]{\left\vert#1\right\rangle}
\newcommand{\bra}[1]{\left\langle#1\right\vert}

\newcommand{\figs}{Figs.~}

\title{Entanglement of magnetic impurities through electron scattering in an electric field}

\author{Oscar Lazo-Arjona$^{1}$, Guillermo Cordourier-Maruri$^{2,3}$ and Romeo de Coss$^{3}$ \\
\\
$^{1}$ Facultad de Ingenier\'ia, Universidad Aut\'onoma de Yucat\'an, \\ 
      A.P. 150 Cordemex, M\'erida, Yucat\'an 97310, M\'exico.\\
\\
$^{2}$ Department of Physics and Astronomy, University College London, \\
       Gower Street, London WC1E 6BT, United Kingdom.\\
\\
$^{3}$ Departamento de F\'isica Aplicada, Centro de Investigaci\'on y de Estudios Avanzados \\
       del Instituto Polit\'ecnico Nacional, A.P. 73 Cordemex, M\'erida, Yucat\'an 97310, M\'exico.\\
}

\begin{document}
\large
\maketitle
\begin{abstract} 
                  \large
                  We show that the entanglement between two distant magnetic impurities, generated via electron scattering, can be easily modulated by
                  controlling the magnitude of an applied external electric field. We assume that the two magnetic impurities are fixed and located on an 
                  one-dimensional quantum wire. A ballistic electron moving through the wire is scattered off by both impurities, so the electron spin can be
                  seen as a mediator between the spins of the impurities. Heisenberg operators are used to describe the interactions between electron and
                  impurities spins. We use a wave guide formalism to model the ballistic electron wave-function. Entanglement control is shown to be possible
                  for three different protocols of entanglement detection. The effect of detection protocols on the entanglement extraction is discussed.
\end{abstract}

\section{Introduction}
\label{intro}

The advantages of quantum information processing (QIP) over its classical counterpart are provided by new quantum resources like entanglement \cite{nielsen}.
Any physical implementation of a QIP has to assure a proper control of entanglement generation between its elements. A special problem arises when the 
elements of a system are located far away to each other. In such situation, a mobile element can be used as a mediator between two far-away elements 
\cite{yama,chen,pop}. The implementation of photons as mobile qubits is widely used, and the quantum information stored in its polarization can be
transported long distances without considerable losses \cite{cirac,yao,tan}. 

In a solid state scenario, another possibility arises: instead of photon polarizations, the electron spin can be used as a mobile qubit \cite{pop}. The Kondo
and RKKY interactions have been shown to be useful to generate entanglement between distant magnetic impurities in an electronic environment \cite{cho1,niz,bayprl12,baync14},
here we study a different approach using a ballistic electron. The electron, as charge carrier, has the advantage of interacting easily with other charges
and being detected without destruction of its quantum information cargo, which is stored in the electron spin. Also, it has been proven the relatively high 
spin decoherence lengths and times present in some material such as GaAs \cite{kika,chen2} and graphene \cite{fucprb12,casrmp09}. Moreover, experimental work
in a new field called quantum electron optics is heading to the control and manipulation of single electrons \cite{fe,hermeline,mcnell}.

The use of ballistic electron spins as mobile qubits, which can interact with fixed spins through scattering processes, has been proposed to implement quantum
logic gates \cite{cor,corprb14}, teleportation \cite{cic}, and entanglement generation \cite{costa,cic2,yuasa,cic3,hida,metprb14}. The entanglement can be
enhanced by the resonance of the electron wave function between the two fixed spins \cite{yuasa}. This behaviour is explained in terms of the
indistinguishability of the possible paths that the electron can follow in the scattering process; the less we know the ballistic electron path, the more
entanglement we can extract.  

In general, this proposal has shown that electron scattering based QIPs have a remarkable resilience, and they require a low level control over the spin interaction \cite{cor}. On the other hand, the involvement of the spatial degrees of freedom is a disadvantage of using electron scattering as an interaction
channel between spins because it makes the scattering a non-unitary operation \cite{cor,cic4}. In order to create a unitary process, we need to implement a
post-selection protocol, such as a spin polarization detection, and the possible implementations will be non-deterministic. The effects of the post-selection
protocol become crucial to the efficiency of the implementation of this proposal.

It is understood that the level of interaction between ballistic and static spins could be controlled by tuning the coupling of ballistic electron and impurity spins, the position of the impurities, and the ratio between kinetic and interaction-devoted energies. The two first options are hard to implement,
specially if the static spins are magnetic impurities. However, the kinetic energy can be easily tuned by applying an external electric field. 

In this work, we show how the entanglement of magnetic impurities created through electron scattering, can be controlled using an external electric field. We
will also show the influence of the electron detection protocol on the impurities entanglement. We analyse three different detection protocols for 
entanglement extraction: i) detecting the electron spin and charge, ii) detecting only the electron charge, and iii) when neither charge nor spin detection
is performed.
     
The present paper is organized as follows. In \sect II, we show our two-impurity scattering model and describe the method used to find the electronic
transmittivity and reflexivity. In \sect III we calculate the entanglement generation when the protocol includes charge and spin detection. The case where only charge detection is available will be shown in \sect IV. In \sect V we depict the case when neither charge nor spin detection is performed. Finally, we
present our conclusions.

\section{A two-impurity scattering model} \label{scattering}

\begin{figure}[ht]
     \begin{center}
     \resizebox{10cm}{!}{\includegraphics{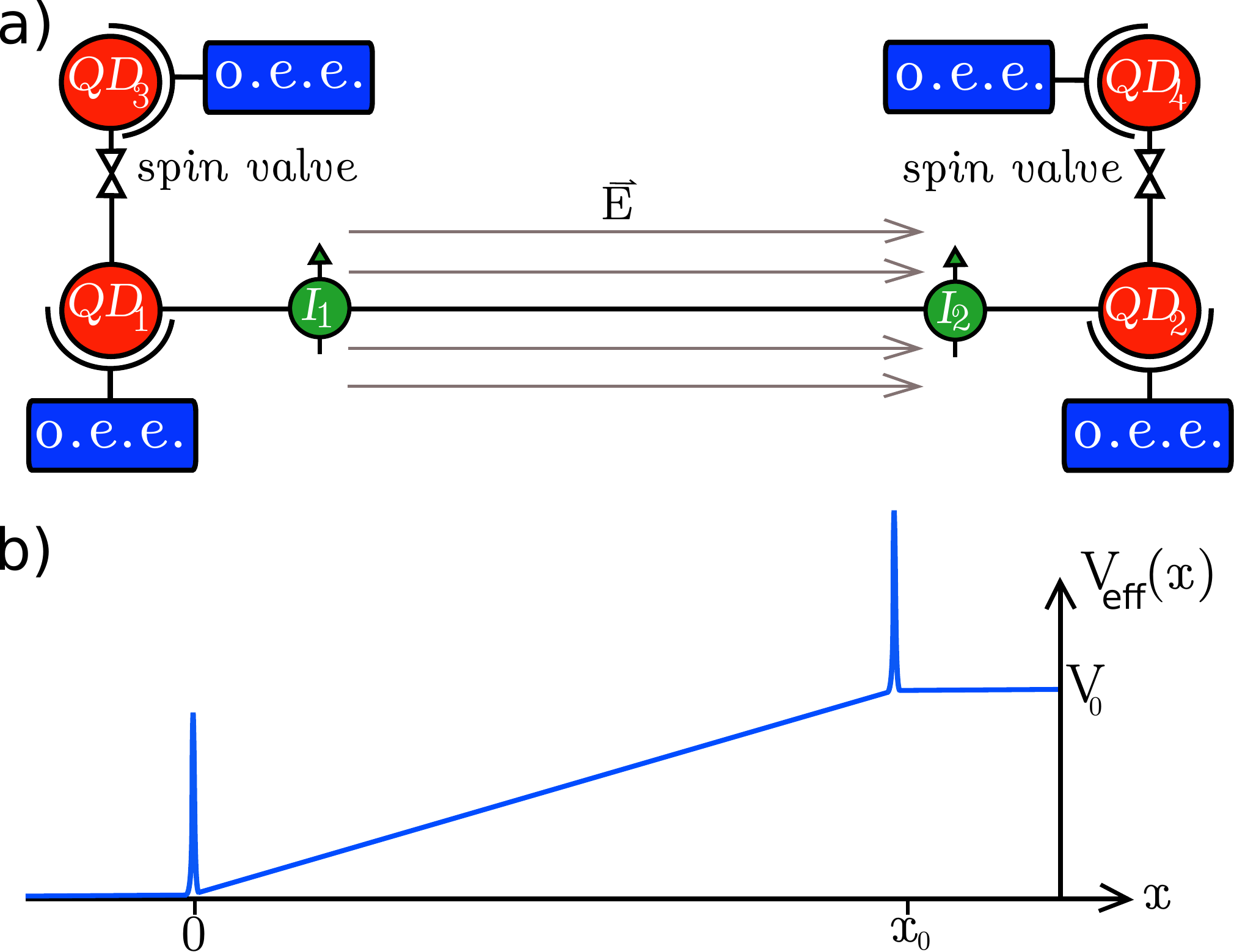}}
     \caption{(Color online) a) Scheme of the device setup. A quantum dot ($QD_1$) acts as a source of a single electron which can be injected through a 
              one-dimensional quantum wire, and be scattered off by two fixed magnetic impurities ($I_1$ and $I_2$), to reach a second quantum dot ($QD_2$)
              acting as a drain. An electric field $\vec{E}$ is established in the zone between the two quantum dots. In order to detect electronic reflection
              and transmission, one-electron electrometers are connected to $QD_1$ and $QD_2$, respectively. The additional quantum dots ($QD_3$ and $QD_4$),
              electrometers and the spin valves can help to perform spin measurements. b) Diagram of the effective potential 
              ($V_{eff}(x) = J\hat{\mbox{\boldmath$S$}}_e\cdot\hat{\mbox{\boldmath$S$}}_1\delta(x) + J\hat{\mbox{\boldmath$S$}}_e\cdot\hat{\mbox{\boldmath$S  
              $}}_2\delta(x-x_0)+V(x)$) depicted by the Hamiltonian of our model.}
     \label{model}
     \end{center}
\end{figure}

We consider a one-dimensional quantum wire where two $1/2$-spin fixed magnetic impurities are located at $x_0$ distance of each other. The quantum wire can 
be physically implemented with carbon nanotubes \cite{white,balents,gunjp06}, graphene nanoribbons \cite{corprb14,tra}, heterostructures \cite{aus} or
quantum Hall edge states \cite{hermeline,mcnell}. We assume that $x_0$ is long enough to neglect any direct interaction between the impurities. From a
quantum dot ($QD_1$ in \fig 1 a)), acting like a single-electron source, a ballistic electron with a controlled energy $\epsilon$ and spin-polarized state is
injected through the quantum wire. The ballistic electron will suffer multiple back scatterings in the region between the two impurities ($0 < x < x_0$),
where we set an uniform electric field $\vec{E}$ along the quantum wire direction. This could be implemented with an electric potential $V_0 = |\vec{E}|x_0$.
Finally, the electron could be transmitted, arriving to a second quantum dot ($QD_2$ in \fig 1 a)), or reflected, returning to the source. Theoretically, it
is possible to re-trap a single electron in a quantum dot after propagation, by controlling the electron kinetic energy through gate potentials, in order to
create bound states \cite{gunjp06}. It has to be considered that there is always a probability that the electron tunnel-out of the quantum dot. However,
experimental works have reported that the efficiency of this process can be over 90 \% \cite{hermeline,mcnell}. The direct spin interaction (electron -
impurity) can be modelled as a short range Heisenberg exchange. If we locate an impurity at $x=0$, the evolution of the system is described by the 
Hamiltonian

\begin{equation}\label{h1}
\hat{H} = \frac{P^2}{2m} + J\hat{\mbox{\boldmath$S$}}_e\cdot\hat{\mbox{\boldmath$S$}}_1\delta(x) 
+ J\hat{\mbox{\boldmath$S$}}_e\cdot\hat{\mbox{\boldmath$S$}}_2\delta(x-x_0) + V(x),
\end{equation}

\noindent
where $P =-i\hbar\frac{d}{dx}$ is the linear momentum operator and $m$ is the effective mass of a ballistic electron in the quantum wire (i.e. in a GaAs
wire, effective mass is 0.067 times the free electron mass). $\hat{\mbox{\boldmath$S$}}_e$ and $\hat{\mbox{\boldmath$S$}}_i$ are the dimensionless spin
operators of the electron and the $i$-th impurity. The coupling factor $J$ (with units of energy times length) is the interaction strength which depends on
the material the impurity is made. 

To avoid some extra resonant behaviour and to simplify the analysis, here we assume that the zone affected by the applied electric field matches with the positions of the magnetic impurities, as shown in \fig 1 b). Thus, the electric potential is

\begin{equation}\label{wfe}
V(x) = \left\{ \begin{array}{rcl} 0  & \mbox{for} & x < 0      \\ 
                   \frac{V_0}{x_0}x  & \mbox{for} & 0 < x < x_0 \\
                                V_0  & \mbox{for} & x_0 < x \end{array}\right.
\end{equation}

\noindent
The linear change in the applied voltage in a nano-structure is justified if we note that our electron reservoirs (the quantum dots) do not have a high number
of modes.  

The ballistic wave function has to consider the spatial and spinorial spaces as 

\begin{equation}\label{wf_tot}
\Psi(x,\zeta) =  \sum_j \psi_j(x)|\zeta\rangle_j
\end{equation}

\noindent
with $\{ \zeta_j \}$ the base of the three particles spinor state and $\psi_j(x)$ the corresponding spatial wave functions. The three-spin system has a total
magnetic momentum number $m_T=$($\pm 3/2$, $\pm 1/2$) which remains constant during the scattering process. To avoid trivial results we set our spinor state
in the $m_T=\pm 1/2$ subspace, allowing a spin flip to occur, which is prohibited in the others subspaces. In this way, we assume that the ballistic electron
spin is always initially in the  $\ket{\downarrow}_e$ state measured on the $z$ axis, and the first and second impurity spins are set in the states 
$\ket{\uparrow}_1$ and $\ket{\uparrow}_2$, respectively. 

An alternative initial state with the anti-parallel impurity spins ($\ket{\uparrow}_1$ and 
$\ket{\downarrow}_2$ or $\ket{\downarrow}_1$ and $\ket{\uparrow}_2$) add an extra asymmetry in the system. Although this asymmetry does not eliminate all the
resonant effects, it restricts the constructive quantum interference \cite{cormplb11}. It also affects the indistinguishability of the Feynmann paths of the
scattered electron, decreasing the level of entanglement created \cite{hida,cic3}. Because our main aim is to analyse the effect of electric field and the
sake of clarity, we consider the impurities initial spin state as always parallel.

Thus, the initial spinor of system is $\ket{\downarrow}_e\ket{\uparrow}_1\ket{\uparrow}_2 = \ket{\downarrow\uparrow\uparrow}$ (In the following
we omit the subscripts), and after the electron scattering the spinor evolves to a superposition of the three basis states of $m_T = 1/2$ subspace 
($\ket{\uparrow\uparrow\downarrow}$, $\ket{\uparrow\downarrow\uparrow}$ and $\ket{\downarrow\uparrow\uparrow}$ which correspond to the subscripts $j=1$, $2$
and $3$ in \eq \ref{wf_tot}). The same analysis can be done for the symmetrical sub-space $m_T = -1/2$.

The spatial wave function is modelled as a plane wave

\begin{equation}\label{wf1}
\psi_j(x) = a_j e^{ikx} + r_j e^{-ikx},
\end{equation}

\noindent
for $x<0$, with a wave number $k = \sqrt{2m\epsilon}/\hbar$ and $a_j$ and $r_j$ the probability amplitudes. In the region $0\leq x \leq x_0$, due to the
applied electric potential $V_0$, the solutions of the Schr\"odinger's equation are the linearly independent {\it Airy} functions ($Ai(x)$, $Bi(x)$)
\cite{allprb94,miyjap98,minjap11,buamr12} as

\begin{eqnarray}\nonumber\label{wf2}
\psi_j(x) & = & b_j Ai\left[\left(\frac{2mx_0^2}{\hbar ^2 V_0^2} \right)^{\frac{1}{3}}\left(\frac{V_0}{x_0}x-\epsilon\right)\right] + 
 \\
& & c_j Bi\left[\left(\frac{2mx_0^2}{\hbar ^2 V_0^2} \right)^{\frac{1}{3}}\left(\frac{V_0}{x_0}x-\epsilon\right)\right],
\end{eqnarray}

\noindent
$b_j$ and $c_j$ are the probability amplitudes. When $x_0 < x$, the transmitted wave function is propagating to the right as a plane wave
 
\begin{equation}\label{wf3}
\psi_j(x) = \tau_j e^{iqx},
\end{equation}

\noindent 
with $q = \sqrt{2m(\epsilon-V_0)}/\hbar$ and $\tau_j$ the transmitted probability amplitude. To obtain the reflection and transmission probability amplitudes
of each spinor state ($r_j$ and $\tau_j$), we set boundary conditions at the points $x=0$ and $x=x_0$ ensuring the continuity of the wave function. The
inclusion of delta potentials in the Hamiltonian will cause a discontinuity in the wave function derivative, which can be obtained evaluating the limits 

\begin{equation}
\lim_{\Delta x \to 0} \int_{-\Delta x}^{\Delta x} \hat{H}\Psi(x)dx =\lim_{\Delta x \to 0} \int_{-\Delta x}^{\Delta x} E \Psi(x)dx,
\end{equation}

\noindent
and

\begin{equation}
\lim_{\Delta x \to 0} \int_{x_0 - \Delta x}^{x_0 + \Delta x} \hat{H}\psi(x)dx =\lim_{\Delta x \to 0} \int_{x_0 - \Delta x}^{x_0 + \Delta x} E \psi(x)dx.
\end{equation}

\noindent
Then, the boundary conditions are

\begin{equation}
\Psi'(0_+,\zeta)-\Psi'(0_-,\zeta)=\frac{2mJ}{\hbar^2}\hat{\mbox{\boldmath$S$}}_e\cdot\hat{\mbox{\boldmath$S$}}_1\Psi(0,\zeta),
\end{equation}
 
\noindent
and

\begin{equation}
\Psi'(x_{0+},\zeta)-\Psi'(x_{0-},\zeta)=\frac{2mJ}{\hbar^2}\hat{\mbox{\boldmath$S$}}_e\cdot\hat{\mbox{\boldmath$S$}}_2\Psi(x_0,\zeta).
\end{equation}

In addition to the boundary conditions, we have the conditions describing the initial spinor state of the system ($a_j$ values). With all this, we can solve
the system of linear equations for $r_j$, $b_j$, $c_j$, $\tau_j$, allowing us to know the state to which the system evolves. The probability amplitudes of 
the stationary wave-function (\eq \ref{wf_tot}) can be normalized using the flux conservation
 
\begin{equation}\label{norm}
\sum_j \left| r_{j} \right|^2 + \frac{q}{k} \sum_j\left| \tau_{j} \right|^2 = \sum_j \left| r_{j} \right|^2 + \sum_j\left| t_{j} \right|^2 = 1,
\end{equation}

\noindent
with $t_j = \sqrt{q/k} \tau_{j}$.

Hence, knowing the values of all the probability amplitudes, we can obtain the density matrix ($\rho$) of the scattered state. From that, we calculate the
transmission and reflection probabilities, and the amount of entanglement generated between the spins of the impurities. To obtain the amount of entanglement
we use the {\it concurrence} value ($C$) defined as \cite{woo}

\begin{equation}\label{c}
C = Max\{0,\lambda_1-\lambda_2-\lambda_3-\lambda_4\},
\end{equation} 

\noindent
where $\lambda_i$ are the eigenvalues in decreasing order of the matrix $\sqrt{\sqrt{\rho}\tilde{\rho}\sqrt{\rho}}$, with

\begin{equation}\label{c}
\tilde{\rho} = (\sigma_y\otimes\sigma_y)\rho^*(\sigma_y\otimes\sigma_y),
\end{equation}

\noindent
$\sigma_y$ is usual Pauli matrix and $\rho^*$ is the complex conjugate of the density matrix. 

\section{Spin and charge detection protocol} \label{spcha}

In the first protocol, we consider that a post-selection of the electron position (charge detection) and spin direction is performed. This can be done with
the help of two additional qudots ($QD_3$ and $QD_4$ in \fig 1 a)) connected with the main qudots through spins valves which filter certain spin
polarization. The occupancy of $QD_3$ and $QD_4$ can be tested with one-electron electrometers, implemented by detecting Coulomb blockade oscillations 
\cite{hermeline,mcnell,fieprl93}. The post-selection will cause further collapse of the wave function into one of four outcomes: either that the electron is
transmitted with the spin changed or unchanged, or that it is reflected with the spin changed or unchanged.

Considering that the initial spinor state is $\ket{\downarrow\uparrow\uparrow}$, if the electron spin is measured to have been left unchanged, then the spin
of the system collapses to $\ket{\downarrow\uparrow\uparrow}$ and the state of the impurities is the separable state $\ket{\uparrow\uparrow}$. However, if
the spin is measured to have changed (and the electron transmitted), the global state becomes 
$\Psi _t (x,\zeta ) = e^{iqx}(t_1\ket{\uparrow\uparrow\downarrow} + t_2\ket{\uparrow\downarrow\uparrow})/\sqrt{P_t}$. Here  $P_t= |t_1|^2+|t_2|^2$, is the
normalization constant associated with the probability of success of the electron being transmitted in an entangled state. Since all information related to
the electron is extracted in the post-selection process, the entanglement of the electron with the impurities is destroyed and the state of the impurities is
purified. In this case, the state of the impurities is described by the reduced density matrix 

\begin{equation}
\rho_{imp}=Tr_{x,e}(\rho)= \frac{1}{P_t} \left(\begin{array}{cccc}
       0  & 0 & 0 & 0\\
       0 & |t_1|^2   & t_1 \bar{t_2}  & 0\\
       0 & t_2 \bar{t_1}   & |t_2|^2  & 0\\
       0 & 0 & 0 & 0
     \end{array}\right),
\end{equation}

\noindent
where $Tr_{x,e}(\rho)$ indicates the partial trace of the complete density matrix $\rho$, over the space of the position and spin of the electron. The concurrence takes the form

\begin{equation}
C_T = \frac{2|t_1 \bar{t_2}|}{|t_1|^2+|t_2|^2}=\frac{2|t_1 \bar{t_2}|}{P_t},
\end{equation}

\noindent
with similar expression for $C_R$, in the case when a reflected electron is detected with its spin changed.

\begin{figure}[ht]
     \begin{center}
     \resizebox{8cm}{!}{\includegraphics{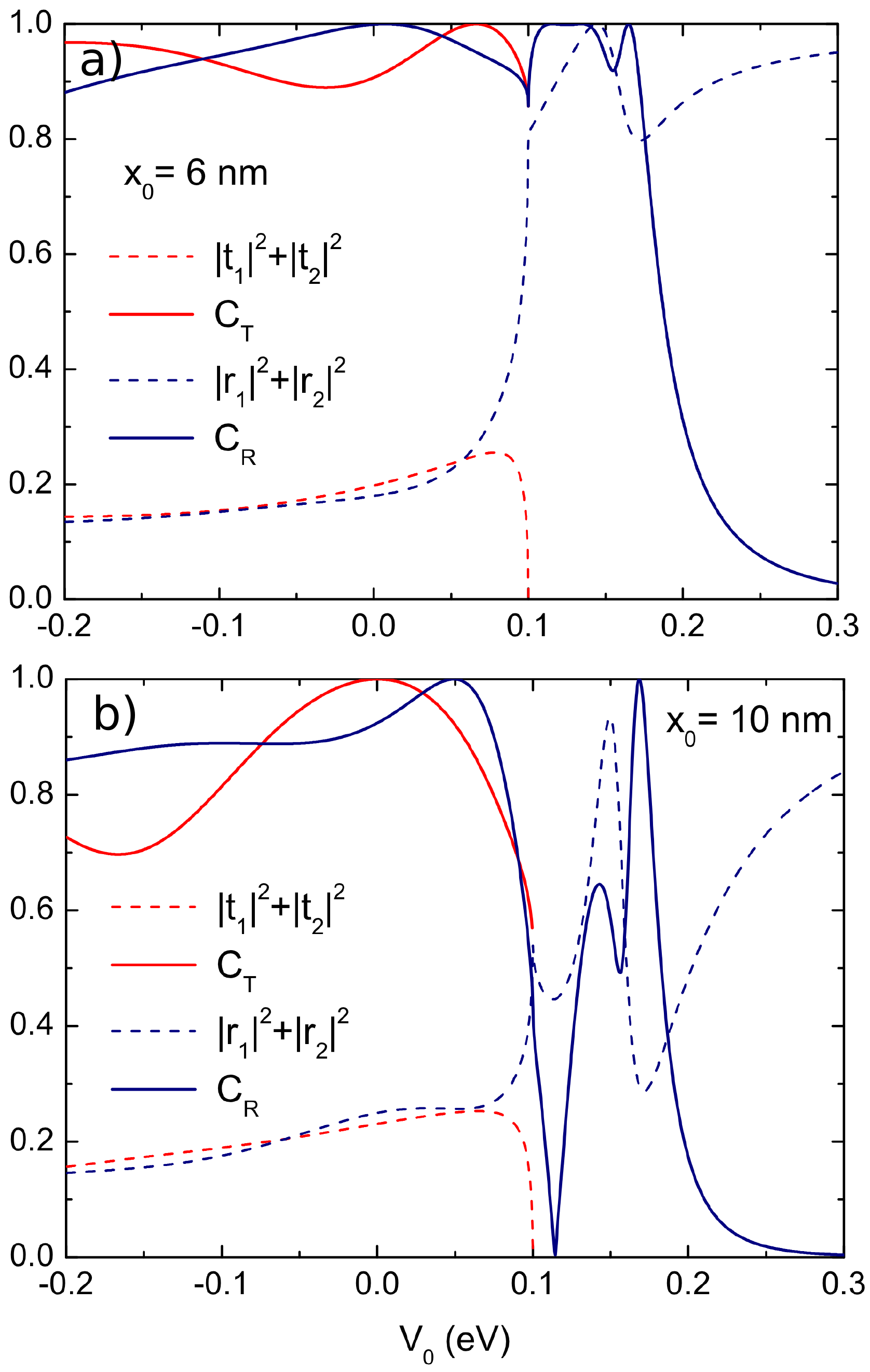}}
     \caption{(Color online) Concurrence for a transmitted (reflected) electron $C_T$ ($C_R$) in red (blue) continuous line, and the probability of success
              in red (blue) dashed line as a function of the applied electric potential $V_0$, following the charge and spin post-selection protocol. The
              magnetic impurities are located on a GaAs quantum wire, with coupling constant $J = 4$ eV\r{A} and ballistic electron energy of $\epsilon = 100$
              meV  for a separation distance of a) $x_0 = 6$ nm and b) $x_0 = 10$ nm.}
     \label{model}
     \end{center}
\end{figure}

\figs 2 and 3 show the concurrence and its probability of success, using the first protocol for a transmitted and reflected electron detection as a
function of the applied potential $V_0$, when the two impurities are separated $x_0 = 6$,$10$, and $100$ nm on a GaAs quantum wire. The ballistic electron
energy is $\epsilon = 100$ meV, and the coupling constant is $J = 4$ eV\r{A}, which is inside the range of electron–impurity spins coupling constants in an
electron scattering process \cite{cic3,bona}. Is important to note that the magnitude of the spins interactions will depend on the ratio between the coupling
factor $J$ and the kinetic energy $\epsilon$.

Note that in all cases with transmission and reflection electron detection, the concurrence can reach the
maximum value $C_{T,R}=1$, modulating the height of the electric potential $V_0$. In the reflection case, $C_R$ can be tuned to reach every value from $0$ to
$1$ for all the separation distances.  

\begin{figure}[ht]
     \begin{center}
     \resizebox{8.1cm}{!}{\includegraphics{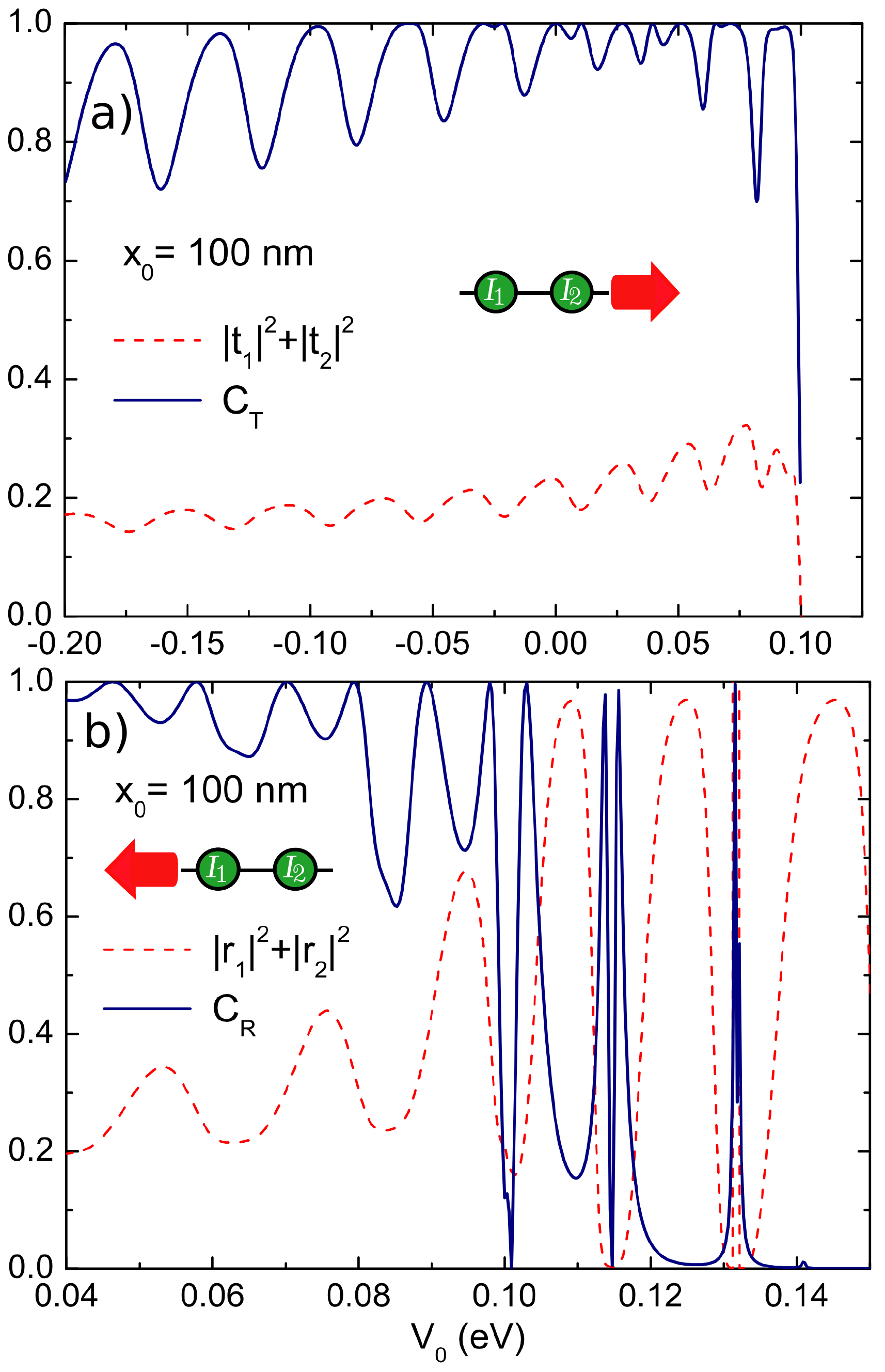}}
     \caption{(Color online) Concurrence for a) transmitted ($C_T$) and b) reflected electron ($C_R$) in blue solid line, and the probability of success in
              red dashed line for both cases as a function of the applied electric potential $V_0$, following the charge and spin post-selection protocol. 
              The magnetic impurities separation is $x_0 = 100$ nm in a GaAs quantum wire, the coupling constant is $J = 4$ eV\r{A} and the ballistic electron
              energy is set to $\epsilon = 100$ meV.}
     \label{model}
     \end{center}
\end{figure}

As expected, the transmission plots are truncated for an electric potential equal to the ballistic electron energy ($V_0 = \epsilon = 100$ meV in the present
examples), and the electron is totally reflected for potentials higher than $\epsilon$. If we continue to increase the $V_0$ value, even larger than 
$\epsilon$, the reflected scattering can still produce perfect entanglement ($C_R = 1$). This effect is due to the penetration of the electron wave-function,
which allows the interaction with the second impurity up to certain value of potential, say $V_{max} = \epsilon + \Delta V$. After this potential value, the
spinorial evolution $\ket{\downarrow\uparrow\uparrow} \rightarrow \ket{\uparrow\uparrow\downarrow}$ is unlikely to happen, producing zero concurrence. The
value of $\Delta V$ depends on the scattered electron energy $\epsilon$, the coupling factor $J$, and the separation between impurities $x_0$. In the case
depicted by the \fig 2, this {\it effective reflection region} extends to $V_{max} \approx 350$ meV ($\Delta V \approx 250$ meV) for $x_0 = 6$ nm, and to
$V_{max} \approx 300$ meV ($\Delta V \approx 200$ meV) for $x_0 = 10$ nm.

\fig 3 shows the case when the separation between impurities is increased to $100$ nm. We note that the resonant behaviour is evidenced by the presence of 
pseudo-bound states between $\delta$-function potentials \cite{cor2}, which are more abundant for larger impurities separations and can enhance the spin
interaction. In both, transmission and reflection plots, the oscillations in the concurrence and probability values are not periodic. The non-periodicity
is more evident when $V_0$ is positive due to the resonances between the income and outcome ballistic electron. For this case ($\epsilon = 100$ meV, $J = 4$
 eV\r{A} and $x_0 = 100$ nm) the effective reflection region extends to $V_{max} \approx 140$ meV ($\Delta V \approx 40$ meV), and inside this region the
concurrence can reach values from $0$ to $1$.

\section{Charge detection protocol} \label{cha}

If a charge measuring device is placed at either lead of the wire, the measurement would provide information about the position of the electron and
collapse the wave function into either $\Psi_t (x,\zeta ) = e^{iqx}(t_1\ket{\uparrow\uparrow\downarrow} +t_2\ket{\uparrow\downarrow\uparrow} +t_3\ket{\downarrow\uparrow\uparrow})/\sqrt{P_t}$ or $\Psi_r (x,\zeta ) = e^{-ikx}(r_1\ket{\uparrow\uparrow\downarrow} +r_2\ket{\uparrow\downarrow\uparrow} + r_3\ket{\downarrow\uparrow\uparrow})/\sqrt{P_r}$, depending on the result of the measurement. Here, $P_t=|t_1|^2+|t_2|^2+|t_3|^2$ and $P_r=|r_1|^2+|r_2|^2+|r_3|^2$
are the normalization constants associated with the events of finding a transmitted and a reflected electron, respectively. The probabilities of this events
are $P_t$ and $P_r$, respectively. A charge measuring device can be implemented with the help of electrometers connected to $QD_1$ and $QD_2$ as is shown in
\fig 1. The spin state of the impurities in the event of transmission is given by the reduced density matrix

\begin{equation}
\rho_{imp}=Tr_{x,e}\rho= \frac{1}{P_t} \left(\begin{array}{cccc}
       |t_3|^2  & 0 & 0 & 0\\
       0 & |t_1|^2   & t_1 \bar{t_2}  & 0\\
       0 & t_2 \bar{t_1}   & |t_2|^2  & 0\\
       0 & 0 & 0 & 0
     \end{array}\right).
\end{equation}

\begin{figure}
     \begin{center}
     \resizebox{8cm}{!}{\includegraphics{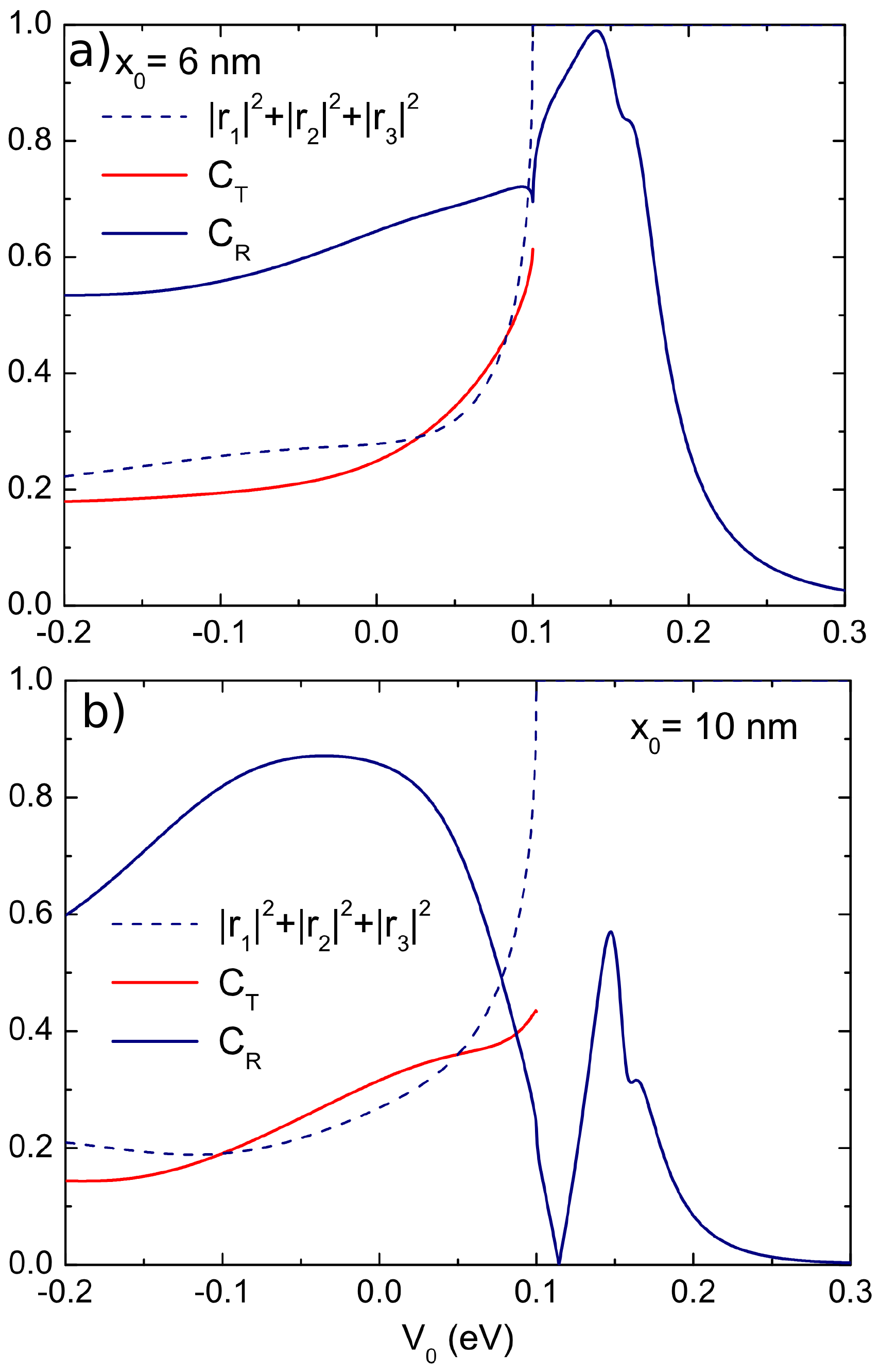}}
     \caption{(Color online) Concurrence for transmitted electron ($C_T$) in red solid, for reflected electron ($C_R$) in blue solid line, and the
              probability electron reflection in dashed line as a function of the applied potential $V_0$, following the charge post-selection protocol. The
              magnetic impurities are located on a GaAs quantum wire, and the separation distance is a) $x_0 = 6$ nm and b) $x_0 = 10$ nm. The coupling
              constant is $J = 4$ eV\r{A} and the ballistic electron energy is $\epsilon = 100$ meV.}
     \label{model}
     \end{center}
\end{figure}

\noindent
In this case, the concurrence has the analytical form

\begin{equation}\label{cp2}
C_T = \frac{2|t_1 \bar{t_2}|}{|t_1|^2+|t_2|^2+|t_3|^2} = \frac{2|t_1 \bar{t_2}|}{P_t},
\end{equation} 

\noindent
with a similar result in the event of reflection. \eq \ref{cp2} implies that the concurrence values could be lower than in the previous protocol, despite a higher value of the probability of success $P_t$. This reduction in the concurrence is due to the existence of quantum correlations and entanglement
between the ballistic electron and the spins of the impurities, which are not destroyed by the charge measurement. 

In \figs 4 and 5, we show the concurrence and the probability of success as a function of the applied potential in the charge detection protocol, for
transmission and reflection events, respectively. The model settings ($\epsilon, J$) are the same as in the previous section. Transmission probability is not
plotted, because in this case transmission and reflection have complementary probability ($|t_1|^2+|t_2|^2+|t_3|^2=1-|r_1|^2+|r_2|^2+|r_3|^2$). The plots
show that when the applied potential is greater than the electron energy the reflection event has a probability of success equal to $1$. 

\begin{figure}
     \begin{center}
     \resizebox{8cm}{!}{\includegraphics{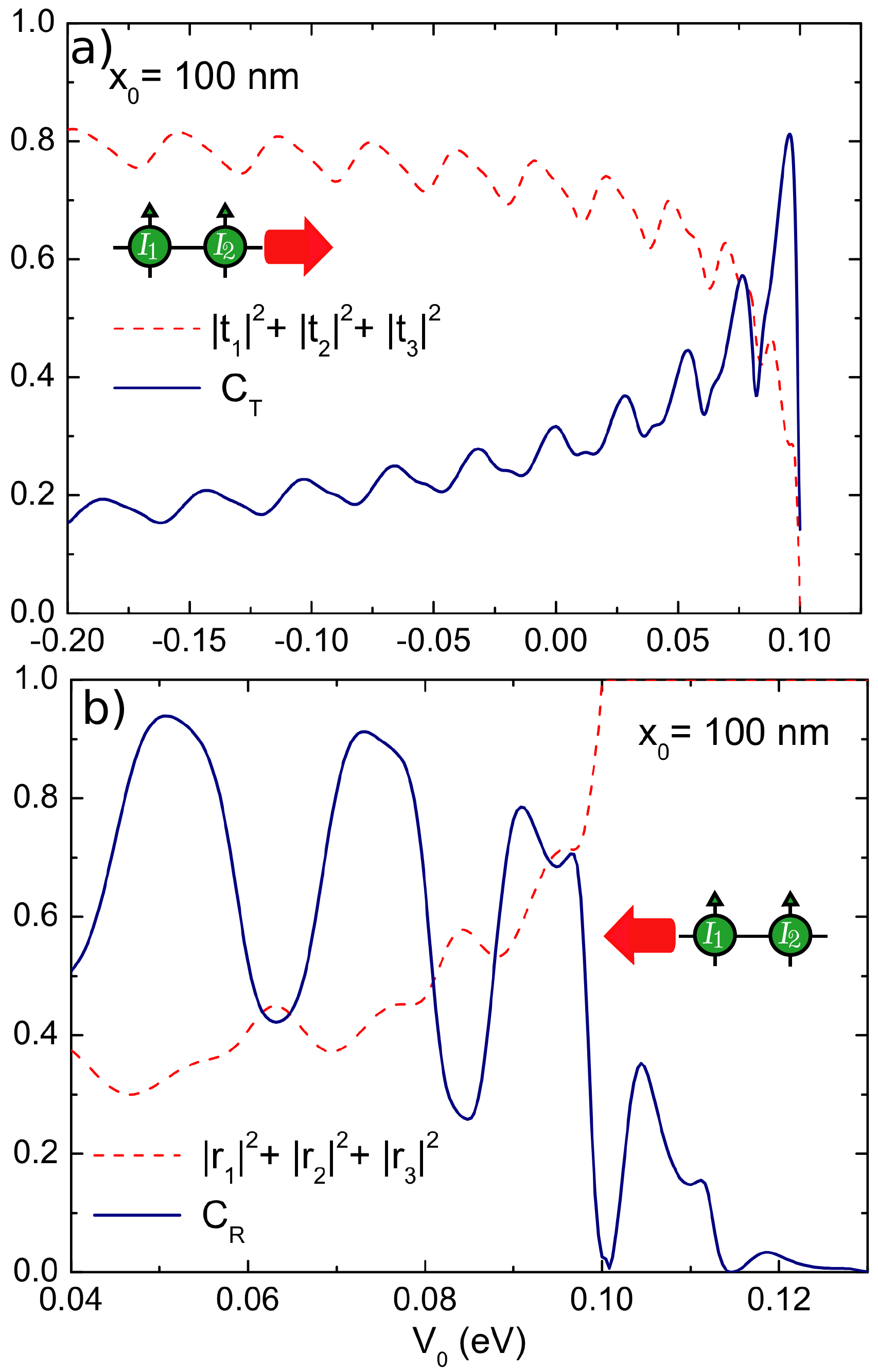}}
     \caption{(Color online) Concurrence for a) transmitted ($C_T$) and b) reflected electron ($C_R$) in blue solid line, and the probability of success
              in red dashed line for both cases, as a function of the applied electric potential $V_0$ and following the charge post-selection protocol. 
              The separation between magnetic impurities is $x_0 = 100$ nm on a GaAs quantum wire, the coupling constant $J = 4$ eV\r{A}  and the ballistic
              electron energy $\epsilon = 100$ meV.}
     \label{model}
     \end{center}
\end{figure}

In \fig 4, the separation between impurities is a) $x_0 = 6$ nm and b) $x_0 = 10$ nm. As expected, the values of concurrence are lower than in the previous
protocol, being considerably higher for the reflection case. It should be noted that for a separation between impurities of $6$ nm, perfect entanglement 
($C_R = 1$) in the effective reflection region can be found. This behaviour is due to the fact that the penetration depth needed to perform the spin
interaction with the second impurity ($\approx 2$ nm), is in the same order of magnitude that the ballistic electron wave-length ($\approx 2.3$ nm for 
$\epsilon = 100$ meV), increasing the possibilities of the interaction due to resonance present in the system.          

In \fig 5, the separation between impurities is increased to $x_0 = 100$ nm, and the resonant behaviour becomes more evident. In this case, the reflection
concurrence is still larger than the transmitted one, but the concurrence in transmission increases as the impurities distance increases due to resonant
effects in the region where $V_0$ is near to $\epsilon$. Note that the effective reflection region does not depend on the type of post-selection protocol,
and it has the same extension that in the previous protocol.

In \figs 4 and 5, a qualitative difference in the behaviour of $C_R$ for $V_0 < \epsilon$ and $V_0 > \epsilon$ is shown. When $V_0 < \epsilon$ the
concurrence value is smooth and rather high. This is because the system is in an diffusion region with a low amount of reflection probability, but we expect that the electron will interact with both impurities, increasing the indistinguishability of the electron path and increasing the entanglement amount. When  $V_0 > \epsilon$ we observe a changing behaviour. This is because the system is in a tunnelling region and the probability of interaction with the second impurity is an evanescent function, but the probability of interaction with the first impurity is oscillatory.

\section{Protocol without spin and charge detection} \label{wd}

In this section, we study the case when neither spin nor charge detection is performed. It is important to point out that in this case, even if we do not know if the electron was transmitted or reflected, we needed to know if the scattering took place, or not. In order to fulfil this requirement, we propose to perform ensemble measurements to obtain information about the efficiency of the scattering process. In these series of experiments, to obtain information about the transmission or reflection result is not necessary. The rate of success of scattering process will depend on the control that we have of the experimental system, and we suppose that it can be perfectible. Then, we can consider that, unless this success rate is close to the unity, this protocol is a non-deterministic process. In the present situation, the final state is given by the normalization condition in \eq \ref{norm},

\begin{equation}
\Psi = \left\{ \begin{array}{rcl}
e^{-ikx}(r_1\ket{\uparrow\uparrow\downarrow} +r_2\ket{\uparrow\downarrow\uparrow} +r_3\ket{\downarrow\uparrow\uparrow}) & \mbox{for} & x < 0      \\
e^{iqx}(t_1\ket{\uparrow\uparrow\downarrow} +t_2\ket{\uparrow\downarrow\uparrow} +t_3\ket{\downarrow\uparrow\uparrow})  & \mbox{for} & x > x_0 \end{array} \right.,
\end{equation}

\noindent
and the final spin state of the impurities is mixed, which can be depicted by the reduced density matrix

\begin{eqnarray}\nonumber\label{den3}
\rho_{imp} & = &Tr_{x,e}(\ket{\Psi}\bra{\Psi})
 \\
& = & \left(\begin{array}{cccc}
       |r_3|^2 + |t_3|^2& 0 & 0 & 0\\
       0 & |r_1|^2 + |t_1|^2 & r_1 \bar{r_2} + t_1 \bar{t_2} & 0\\
       0 & r_2 \bar{r_1} + t_2 \bar{t_1} & |r_2|^2 + |t_2|^2 & 0\\
       0 & 0 & 0 & 0
     \end{array}\right).
\end{eqnarray}

\noindent 
Using \eq \ref{den3}, the concurrence has the following analytical form 

\begin{equation}
C = 2 \left| r_1 \bar{r_2} + t_1 \bar{t_2} \right|,
\end{equation} 

\noindent
for $\epsilon > V_0$, and $C = 2 \left| r_1 \bar{r_2} \right|$ for $\epsilon < V_0$. In this protocol we expect to have a lower amount of entanglement, but
with the advantage of certainty of the entanglement success.  

It should be noted that in the present and in previous protocol, when the final electron spin polarization is left unknown, the quantum correlation related
to the electron spin can be considered as a source of decoherence, affecting the impurities system.

In \fig 6, we show how the concurrence changes with the applied potential $V_0$, for impurities separation of $6$, $10$, and $100$ nm. Note that the resonant
behaviour becomes predominant as the separation between impurities increases. In the region where $V_0 < \epsilon$, the amount of entanglement is similar in
all cases. A major difference is present in the effective reflection region ($100$ meV $\le V_0 \le$ $250$ meV approximately for $x_0 = 6$ nm, $100$ meV
 $\le V_0 \le$ $200$ meV approximately for $x_0 = 10$ nm, and $100$ meV $\le V_0 \le$ $140$ meV approximately for $x_0 = 100$ nm), where merely controlling
the potential allows us to obtain total entanglement as well as any concurrence from $0$ to $1$, for a separation between impurities of $6$ nm. The other two
separation distances present a regular amount of generated entanglement because in these cases, the penetration depth needed to generate a proper interaction with the second impurity is larger than the ballistic electron wave-length, as we explained in the previous section.

\begin{figure}[h]
     \begin{center}
     \resizebox{8cm}{!}{\includegraphics{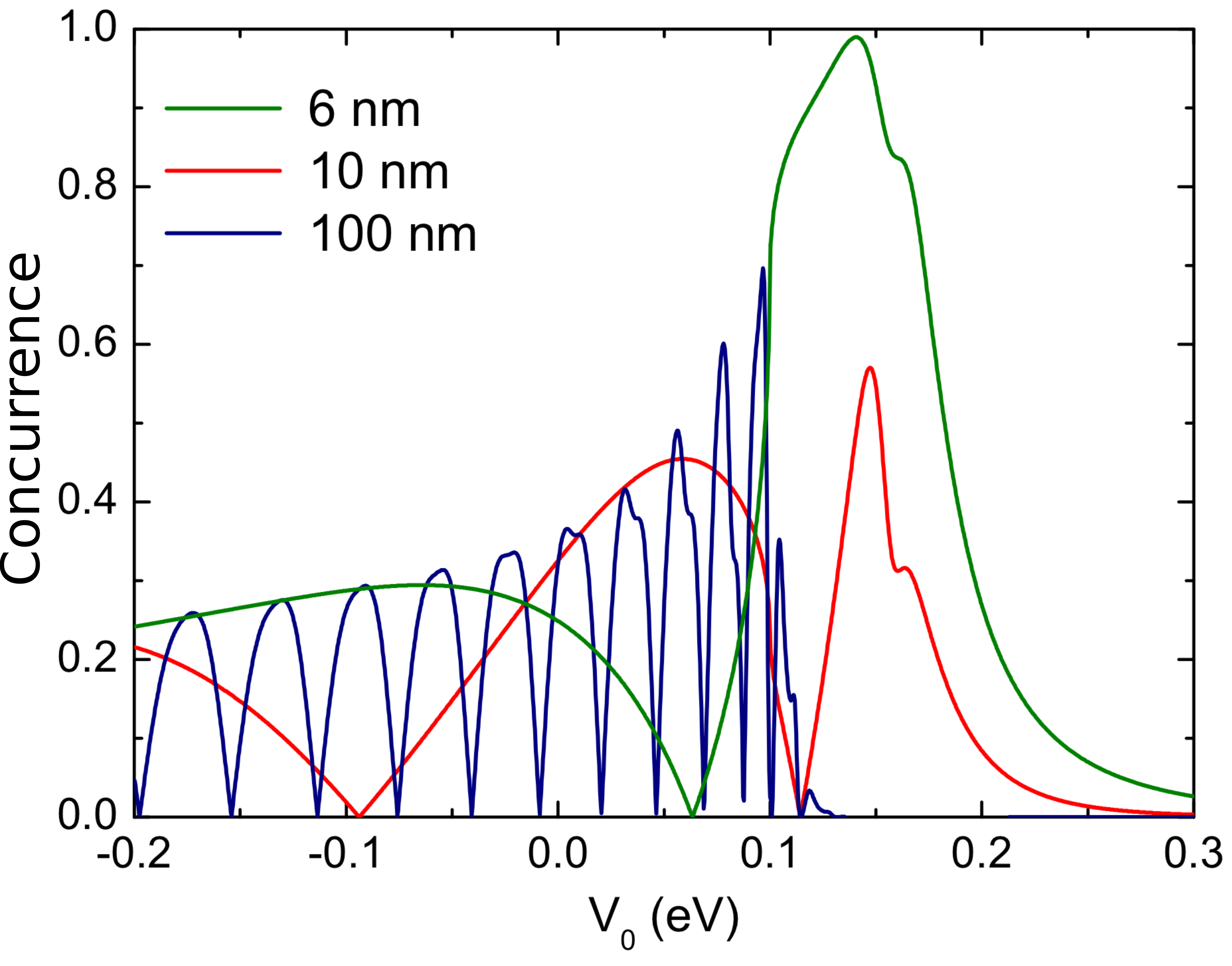}}
     \caption{(Color online) Concurrence $C$ as a function of the applied potential $V_0$, following the protocol without spin and charge detection for a
              separation between magnetic impurities of $x_0 = 6$ nm (green line), $x_0 = 10$ nm (red line), and $x_0 = 100$ nm (blue line). The coupling
              constant and the ballistic electron energy are set to $J = 4$ eV\r{A} and $\epsilon = 100$ meV, respectively.}
     \label{model}
     \end{center}
\end{figure}

In \fig 6, the dependence of the effective reflection region width ($\Delta V$) on the separation between impurities becomes more evident: $\Delta V$
decreases as the separation between the impurities is increased. This effect can be explained from the fact that increasing the potential slope in a 
system with a large separation between impurities, a larger penetration depth will be needed to establish the interaction with the second impurity.  

Note that the perfect entanglement generated in $x_0 = 6$ nm (at $V_0 \approx 140$ meV) is present in the three protocols, implying that the entanglement
between the ballistic electron and the spins of the impurities is null for that potential value, and a pure entangled state can be obtained using any
protocol in a deterministic form.

\section{Conclusions} \label{conclusion}

We show that the proposed system can be used to obtain any degree of entanglement between the impurity spins by merely controlling the electric potential
applied between the impurities on a quantum wire. This can be done in a variety of configurations of the system, say separation between impurities ($x_0$),
spins coupling factor ($J$), and ballistic electron energy ($\epsilon$). The entanglement control was shown to be possible for three different protocols of
entanglement detection. A resonant behaviour, which becomes more evident for larger separation between impurities ($x_0 = 100$ nm), is present in all the
analysed protocols. Resonance enhances the level of spin interaction and can be useful to increase the entanglement between spins of the impurities.

The charge and spin detection protocol demands more control on the system, however, allows the generation of a pure entangled state of the impurities spins.
The probabilities of success are smaller than the probabilities found in the only-charge detection and no spin and charge detection protocols, but high
amounts of concurrence and detection success are found in an {\it effective reflection region}. We defined the effective reflection region as the potential
gap between the value $V_0 = \epsilon$, where the electron transmission is suppressed, and the value $V_0 = \epsilon + \Delta V$, where the interaction with
the second impurity is unlikely to happen. $\Delta V$ depends on the values of $\epsilon$, $J$, and $x_0$ parameters.   

The use of charge detection can increase the probability of success and under certain parameters conditions, high values of concurrence can be found, especially for electron reflection. The no-detection protocol gives certainty of the entanglement success. These two protocols do not suppress the quantum
correlation between the ballistic electron spin and the impurities spins. Nevertheless, we found conditions (e.g. $J=4$ eV\r{A}, $\epsilon = 100$ meV, 
$x_0 = 6$ nm and $V_0 = 140$ meV) where the entanglement generated is perfect ($C=1$) and the low-control no-detection protocol produces a pure entangled
state.  

\section*{Acknowledgements}

GCM acknowledge to Consejo Nacional de Ciencia y Tecnolog\'ia (Conacyt, M\'exico) for a postdoctoral grant. This research was partially funded by Conacyt,
M\'exico, Under Grant No. 83604.


\begin{thebibliography}{}


   \bibitem{nielsen} Nielsen, M. A. and Chuang, I. L., Quantum Computation and Quantum Information. Cambridge University Press, Cambridge, UK, (2000)

   \bibitem{yama} Yamamoto, M., Takada, S., Baeuerle, C., Watanabe, K., Wieck, A. D., and Tarucha, S.: Electrical control of a solid-state flying qubit Nat.
                  Nanotechnol., {\bf 7}, 247 (2012)

   \bibitem{chen} Chen, G. Y., Lambert, N., Chou, C. H., Chen, Y. N., and Nori, F.: Surface plasmons in a metal nanowire coupled to colloidal quantum dots:
                  Scattering properties and quantum entanglement, Phys. Rev. B, {\bf 84}, 045310 (2011)

   \bibitem{pop} Popescu, A. E. and Ionicioiu, R.: All-electrical quantum computation with mobile spin qubits, Phys. Rev. B, {\bf 69}, 245422 (2004)

   \bibitem{cirac} Cirac, J. I., Zoller, P., Kimble, H. J., and Mabuchi, H.: Quantum state transfer and entanglement distribution among distant nodes in a
                   quantum network, Phys. Rev. Lett., {\bf 78}, 3221 (1997)

   \bibitem{yao} Yao, W., Liu, R. B., and Sham, L. J.: Theory of control of the spin-photon interface for quantum networks, Phys. Rev. Lett., {\bf 95}, 
                 030504 (2005)
 
   \bibitem{tan} Tanzilli, S., Tittel, W., Halder, M., Alibart, O., Baldi, P., Gisin, N., and Zbinden, H.: A photonic quantum information interface, Nature,
                {\bf 437}, 116 (2005)

   \bibitem{cho1} Cho, S. Y. and McKenzie, R. H.: Quantum entanglement in the two-impurity Kondo model, Phys. Rev. A, {\bf 73}, 012109 (2006)

   \bibitem{niz} M. Nizama, D. Frustaglia, and Hallberg K.: Quantum correlations in nanostructured two-impurity Kondo systems, Phys. Rev. B, {\bf 86}, 
                 075413 (2012)

   \bibitem{bayprl12} Bayat, A., Bose, S., Sodano, P., and Johannesson, H.: Entanglement probe of two-impurity Kondo physics in a spin chain, Phys. Rev.
                      Lett. {\bf 109}, 066403 (2012).

   \bibitem{baync14} Bayat, A., Johannesson, H., Bose, S., and Sodano.: An order parameter for impurity systems at quantum criticality, Nat. Commun. {\bf 5},
                     3784 (2014)

   \bibitem{kika} Kikkawa, J. M., and Awschalom, D. D.: Lateral drag of spin coherence in gallium arsenide, Nature, {\bf 397}, 139 (1999)

   \bibitem{chen2} Chen, W., Shen, R., Wang, Z. D., Sheng, L., Wang, B. G., and Xing, D. Y.: Quantitatively probing two-electron entanglement with a
                   spintronic quantum eraser, Phys. Rev. B, {\bf 87}, 155308 (2013)

   \bibitem{fucprb12} Fuchs M., Rychkov V., and Trauzettel B.: Spin decoherence in graphene quantum dots due to hyperfine interaction, Phys. Rev. B, 
                     {\bf 86}, 085301 (2012)

   \bibitem{casrmp09} Castro Neto, A. H., Guinea, F., Peres, N. M. R., Novoselov, K. S., and Geim, A. K.: The electronic properties of graphene, Rev. Mod.
                      Phys., {\bf 81}, 109 (2009)

   \bibitem{fe} Feve, G., Mahe, A., Berroir, J. M., Kontos, T., Placais, B., Glattli, D. C., Cavanna, A., Etienne, B., and Jin, Y.: An on-demand 
                coherent single-electron source, Science, {\bf 316}, 1169 (2007)

   \bibitem{hermeline} Hermelin, S., Takada, S., Yamamoto, M., Tarucha, S., Wieck, A. D., Saminadayar, L., Bauerle1, C., and Meunier, T: Electrons surfing 
                       on a sound wave as a platform for quantum optics with flying electrons, Nature, {\bf 477}, 435 (2011) 

   \bibitem{mcnell} McNeil, R. P. G., Kataoka, M., Ford, C. J. B., Barnes, C. H. W., Anderson, D., Jones, G. A. C., Farrer, I., and Ritchie1, D. A.: 
                    On-demand single-electron transfer between distant quantum dots, Nature, {\bf 477}, 439 (2011)

   \bibitem{cor} Cordourier-Maruri, G., Ciccarello, F., Omar, Y., Zarcone, M., de Coss, R., and Bose, S.: Implementing quantum gates through scattering
                 between a static and a flying qubit, Phys. Rev. A, {\bf 82}, 052313 (2010)

   \bibitem{corprb14} Cordourier-Maruri G., Omar, Y., de Coss, R., and Bose, S.: Graphene-enabled low-control quantum gates between static and mobile spins,
                      Phys. Rev. B, {\bf 89}, 075426 (2014)

   \bibitem{cic} Ciccarello, F., Bose, S., and Zarcone, M.: Teleportation between distant qudits via scattering of mobile qubits Phys. Rev. A, {\bf 81},
                 042318 (2010)

   \bibitem{costa} Costa Jr., A. T., Bose, S., and Omar, Y.: Entanglement of two impurities through electron scattering, Phys. Rev. Lett., {\bf 96}, 230501
                  (2006)

   \bibitem{cic2} Ciccarello F., Palma G. M., Zarcone, M., Omar, Y., and Vieira, V. R.: Entanglement controlled single-electron transmittivity, New J. Phys.,
                  {\bf 8}, 214 (2006)

   \bibitem{yuasa} Yuasa, K., and Nakazato, H.: Resonant scattering can enhance the degree of entanglement, J. Phys. A: Math. Theor. {\bf 40}, 297 (2007)

   \bibitem{cic3} Ciccarello, F., Palma, G. M., Zarcone, M., Omar, Y., and Vieira, V. R.: Electron Fabry-Perot interferometer with two entangled magnetic
                  impurities, J. Phys. A: Math. Theor., {\bf 40}, 7993 (2007)

   \bibitem{hida} Hida, Y., Nakazato, H., Yuasa, K., and Omar, Y.: Entanglement generation by qubit scattering in three dimensions, Phys. Rev. A, {\bf 80},
                  012310 (2009)

   \bibitem{metprb14} Metavitsiadis, A., Dillenschneider, R., and Eggert, S.: Impurity entanglement through electron scattering in a magnetic field, Phys.
                      Rev. B, {\bf 89}, 155406 (2014)

   \bibitem{cic4} Ciccarello, F., Browne, D. E., Kwek, L. C., Schomerus, H., Zarcone, M., and Bose, S.: Quasideterministic realization of a universal quantum
                  gate in a single scattering process, Phys. Rev. A, {\bf 85}, 050305(R) (2012) 

   \bibitem{white} White, C. T., and Todorov, T. N.: Carbon nanotubes as long ballistic conductors, Nature, {\bf 393}, 240 (1998)

   \bibitem{balents} Balents, L., and Egger, R.: Spin transport in interacting quantum wires and carbon nanotubes, Phys. Rev. Lett., {\bf 85}, 3464 (2000)

   \bibitem{gunjp06} Gunlycke, D., Jefferson, J. H., Rejec, T., Ramsak, A., Pettifor, D. G., and Briggs G. A.: Entanglement between static and flying qubits
                     in a semiconducting carbon nanotube, J. Phys.: Condens. Matter, 18, 851 (2006)

   \bibitem{tra} Trauzettel, B., Bulaev, D. V., Loss, D., and Burkard, G.: Spin qubits in graphene quantum dots, Nature Phys., {\bf 3}, 192 (2007)

   \bibitem{aus} Auslaender, O. M., Yacoby, A., de Picciotto, R., Baldwin, K. w., Pfeiffer, L. N., and West, K. W.: Experimental evidence for resonant
                 tunneling in a Luttinger liquid, Phys. Rev. Lett., {\bf 84}, 1764 (2000)

   \bibitem{cormplb11} Cordourier-Maruri, G., de Coss, R., and Gupta, V.: Transmission properties of the one-dimensional array of delta potentials, Mod.
                       Phys. Lett. B, {\bf 25}, 1349 (2011)

   \bibitem{allprb94} Allen, S. S., and Richardson, S. L.: Theoretical investigations of resonant-tunneling in asymmetric multibarrier semiconductor 
                      heterostructures in an applied constant electric-field.

   \bibitem{miyjap98} Miyamoto, K., and Yamamoto, H.: Resonant tunneling in asymmetrical double-barrier structures under an applied electric field, J. App.
                      Phys., {\bf 84}, 311 (1998)

   \bibitem{minjap11} Ming, Y., Gong, J., and Zhang, R. Q.: Spin-polarized transport through ZnMnSe/ZnSe/ZnBeSe heterostructures, J. App. Phys., {\bf 110},
                      093717 (2011)

   \bibitem{buamr12} Bu, X., Wang, J., Shi, J., and Zhao, H.: Research on the electronic tunneling in asymmetric dual-quantum-well, Adv. Mater. Res., 
                     {\bf 542}, 953 (2014)

   \bibitem{woo} Wootters, W. K.: Entanglement of formation of an arbitrary state of two qubits, Phys. Rev. Lett., {\bf 80}, 2245 (1998)

   \bibitem{fieprl93} Field, M., Smith, C. G., Pepper, M., Ritchie, D. A., Frost, J. E. F., Jones, G. A. C., and Hasko, D. G.: Measurements of coulomb
                      blockade with noninvasive voltage probe, Phys. Rev. Lett., {\bf 70}, 1311 (1993)

   \bibitem{bona} Meier, f., Bona, G. L., and H\"ufner, S.:Experimental-determination of exchange constants by spin-polarized photoemission, Phys. Rev. Lett.,
                  {\bf 52}, 1152 (1984)

   \bibitem{cor2} Cordourier-Maruri, G.,  Gupta, V., and de Coss, R.: Recurrence in resonant transmission of the one-dimensional array of delta potetials,
                  Mod. Phys. Lett. B, {\bf 28}, 1450016 (2014)


\end{thebibliography}


\end{document}